\documentclass{PoS}

\usepackage[english]{babel}
\usepackage[utf8]{inputenc}
\usepackage[T1]{fontenc}

\usepackage{tikz}
\usepackage{nicefrac}
\usepackage{graphicx}
\usetikzlibrary{decorations.pathmorphing} 
\usetikzlibrary{decorations.pathreplacing}
\usetikzlibrary{decorations.markings}
\usetikzlibrary{arrows}

\newcommand{\dd}{\,\mathrm{d}}
\newcommand{\ee}{\,\mathrm{e}}

\definecolor{JGURed}{RGB}{193,0,42}

\title{Exploring the phase diagram of QCD with complex Langevin simulations}

\ShortTitle{Exploring the phase diagram of QCD with complex Langevin
simulations}

\author{Gert Aarts$^{1}$, Felipe
Attanasio$^{1,2}$, \speaker{Benjamin Jäger}$^{1}$, Erhard Seiler$^{3}$, Dénes
Sexty$^{4,5}$, Ion-Olimpiu Stamatescu$^{4}$.\\
$^{1}$Department of Physics, College of Science, Swansea University, Swansea, 
UK\\
$^{2}$CAPES Foundation, Ministry of Education of Brazil, Brasília, Brazil\\
$^{3}$Max-Planck-Institut für Physik (Werner-Heisenberg-Institut), München,
Germany\\
$^{4}$Institut für
Theoretische Physik, Universität Heidelberg, Heidelberg, Germany\\
$^{5}$Department of Physics, Bergische Universität Wuppertal, Wuppertal,
Germany\\
E-mail: 
\email{B.Jaeger@swansea.ac.uk}, 
}

\abstract{
Simulations of QCD with a finite chemical potential typically lead
to a severe sign problem, prohibiting any standard Monte Carlo
approach. Complex Langevin simulations provide an alternative to sample path
integrals with oscillating weight factors and therefore potentially enable the
determination of the phase diagram of QCD. Here we present results for QCD in
the limit of heavy quarks and show evidence that the phase diagram can be mapped
out by direct simulation. We apply adaptive step-size scaling
and adaptive gauge cooling to ensure the convergence of these simulations.}

\FullConference{The 32nd International Symposium on Lattice Field Theory\\
                 June 23 – 28,  2014\\
                 Columbia University, New York, USA}

\begin{document}

\section{Introduction}

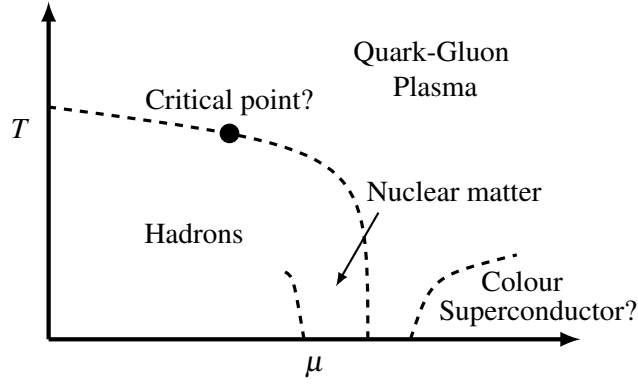
\begin{figure}[h!]
\centering
\begin{tikzpicture}[scale=1.4]
\begin{scope}[>=latex]
\draw[black,->, ultra thick] (0.0,0.0) -- (0.0,3.20); 
\draw[black,->, ultra thick] (0.0,0.0) -- (5.0,0.00); 
\draw[black,very thick, dashed]  (3.0,0.0) .. controls (3,1.80) and (3,1.80) ..
(0,2.2);  
\draw[black,very thick, dashed]  (2.4,0.0) .. controls (2.3,0.60) and (2.3,0.60)
..
(2.2,0.65); 
\draw[black,very thick, dashed]  (3.4,0.0) .. controls (3.6,0.60) and
(3.6,0.60) ..(4.4,0.8); 
\node[below=0.1cm] at (2.5,0) {$\mu$};
\node[left=0.1cm] at (0,2) {$T$};
\node[left=0.1cm] at (2.0,1.0) {Hadrons};
\node[left=0.1cm] at (4.5,2.7) {Quark-Gluon};
\node[left=0.1cm] at (4.2,2.4) {Plasma};
\node[left=0.1cm] at (4.8,1.4) {Nuclear matter};
\node[left=0.1cm] at (5.0,0.55) {Colour};
\node[left=0.1cm] at (5.7,0.25) {Superconductor?};
\draw[black,->, thick] (3.1,1.2) -- (2.7,0.5); 
\filldraw [black] (1.7,1.95) circle (2.5pt);
\node[above=0.1cm] at (1.7,1.95) {Critical point?};
\end{scope} 
\end{tikzpicture}
\caption{A possible sketch of the QCD phase diagram.}
\label{Bild1}
\end{figure}

The current knowledge of the phase diagram of QCD is very limited and a
possible scenario is sketched in figure~\ref{Bild1}. 
The structure of the phase diagram is very important for a 
wide variety of phenomena: the evolution of the early universe, neutron stars 
and heavy-ion collision experiments. Predictions from first principles
 are a clearly desirable goal, but unfortunately direct lattice
simulations of QCD are not possible due to the famous \emph{sign
problem}~[1]: Importance Sampling based Monte Carlo methods cannot be applied to path integrals with a highly oscillating
complex weight. For non-vanishing $\mu$, the fermion determinant
\begin{equation}
\big[\det D(\mu)\big]^\ast = \det D(-\mu^\ast) \, \rightarrow
\, \det D(\mu\neq0) = \left| \det D\right|
\ee^{i \Theta} \in \mathbb{C},
\end{equation} 
adds a complex phase to the path integral,
\begin{equation}
\langle A \rangle = \frac{1}{Z} \int \mathrm{D} U\,  A(U) \, \left|\det D\right|
\textcolor{JGURed}{\ee^{i \Theta}}\, \ee^{-S_G(U)}. 
\end{equation}
The \emph{complex Langevin method} has been shown to provide an alternative to
perform lattice simulation with a non-vanishing chemical potential~[2-12].

\section{Setup}

\begin{figure}[h!] 
	\centering
	\hspace{-0.5cm}
	\begin{minipage}{0.48\linewidth}
	\centering
	\includegraphics[width=\linewidth]{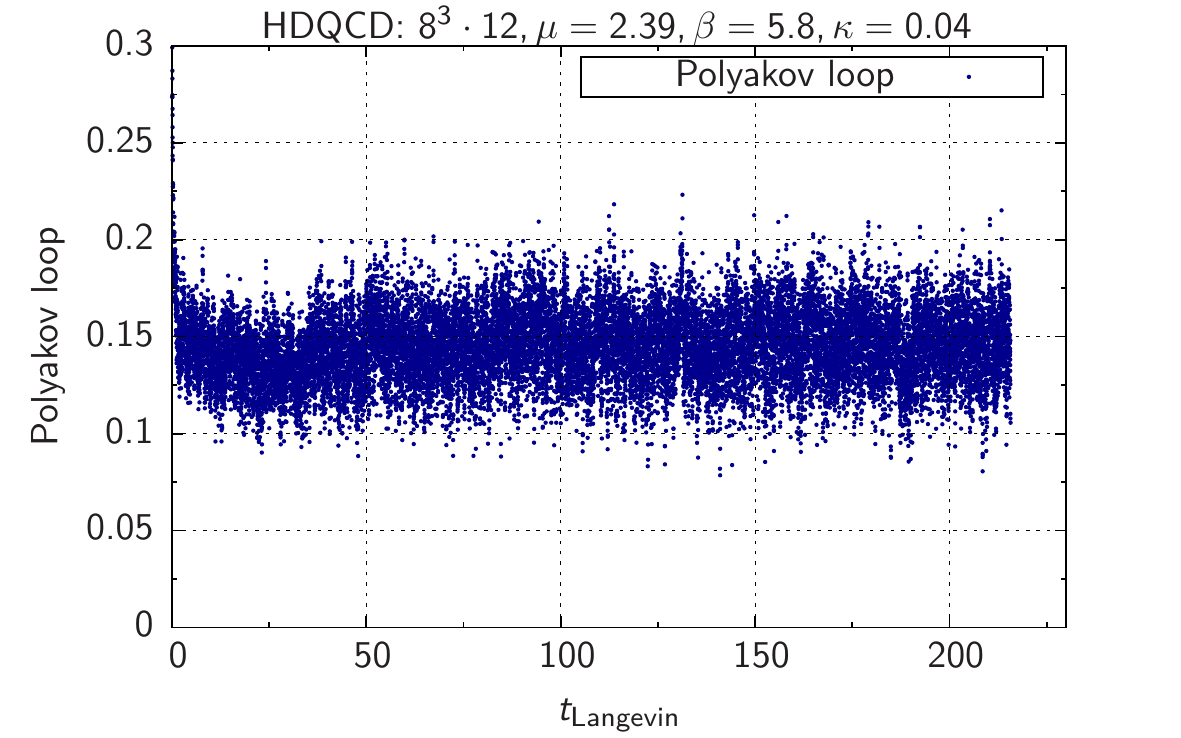}
	\end{minipage}
	\begin{minipage}{0.48\linewidth}
	\centering
	\includegraphics[width=\linewidth]{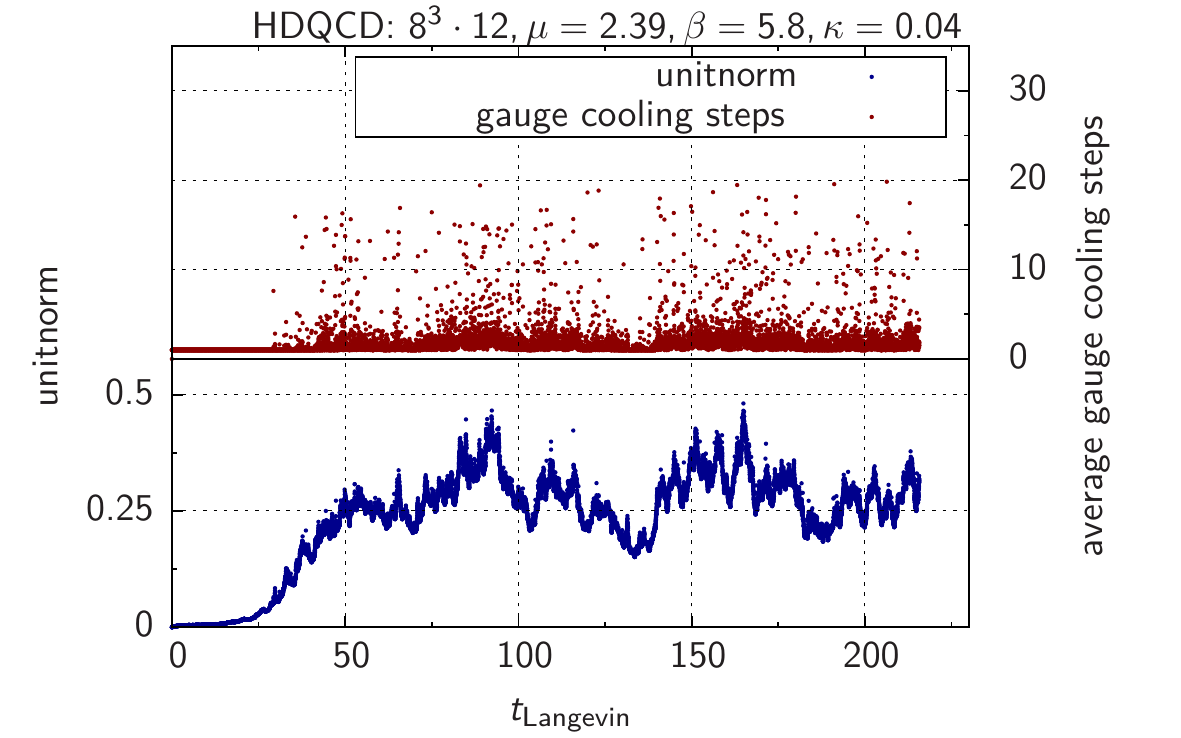}
	\end{minipage}
	
\caption{Left: The real part of the Polyakov loop $P$ as function of the
Langevin time $t$. Right: The unitarity norm, i.e. the distance of the gauge links from the
SU$(3)$ manifold (below) and the average number of gauge cooling steps in a
small Langevin time interval used between each Langevin update (above) as
a function of Langevin time.}
	\label{Bild2}
\end{figure}

The complex Langevin method is based on stochastic
quantization, in which the degrees of freedom $U$, i.e. the gauge links, are
evolved with respect to the so-called Langevin time $t$, suitably discretised,
\begin{equation}
U(t+\epsilon) = R(t)\, U(t) \mathrm{\quad with \quad} R(t) =
\mathrm{exp}\left[ i \lambda \left(- \epsilon \, \partial S + \sqrt{\epsilon}\, \eta  \right)
\right],
\end{equation}
where $\eta$ is Gaussian white noise and $\lambda$ are the Gell-Mann matrices.
These gauge fields are complexified by extending the gauge group
from $\mathrm{SU}(3)$ to $\mathrm{SL}(3,\mathbb{C})$. With these definitions the
expectation value of an observable $A$ can be obtained by integrating 
along the Langevin evolution
\begin{equation}
\langle A \rangle = \frac{1}{T} \int \limits_{0}^{T} A  \left[U(t)\right]\, \dd
t.
\end{equation}
We apply adaptive step-size scaling~[6] and adaptive \emph{gauge cooling}~[7-9]
to avoid excursions into the non-compact "imaginary" directions of
the larger gauge group $\mathrm{SL}(3,\mathbb{C})$. These excursions
are typically generated by numerical artefacts and round-off errors. Convergence
of the Langevin method can be proven, if the distribution in the "imaginary" directions
shows a strong enough falloff and the action is holomorphic~[10,11]. Therefore
we monitor the distribution of our results as well as the unitarity norm, which
is a measure of the distance of the gauge links from the SU$(3)$ manifold and
given by
\begin{equation}
\mathrm{unitnorm} = \mathrm{Tr} \left( U U^{\dagger} - \mathbb{I} \right)^2 \geq
0.
\end{equation}
In the following we present results for QCD in the limit
of heavy quarks using the Wilson formulation. The quarks are considered
static in this approximation, so that the spatial hopping part is neglected. 
This approach is known as heavy dense QCD or HDQCD. In this limit the
fermion determinant simplifies and can be written 
in terms of the (conjugate) Polyakov loops $\mathcal{P}_{\vec{x}}$ and $\mathcal{P}^{-1}_{\vec{x}}$ as
\begin{equation}
\det D(\mu) = \prod_{\vec{x}} \det \left( 1 + h\,\ee^{\mu/T} \,
\mathcal{P}_{\vec{x}} \right)^{2}\det \left( 1 + h\,\ee^{-\mu/T}\,
\mathcal{P}^{-1}_{\vec{x}} \right)^{2},
\end{equation}
where $h = \left( 2 \, \kappa \right)^{N_\tau}$.
For the gluonic part we use the full Wilson gauge action.
Figure~\ref{Bild2} shows the real part of the Polyakov loop $P_{\vec{x}} =
\mathrm{Tr}\,\,\mathcal{P}_{\vec{x}} / 3$ and the unitarity norm as a function
of the Langevin time $t$ for a particular simulation ($8^3\cdot 12$,
$\mu=2.39$, $\beta=5.8$, $\kappa = 0.04$).

\section{Strategy and simulations details}

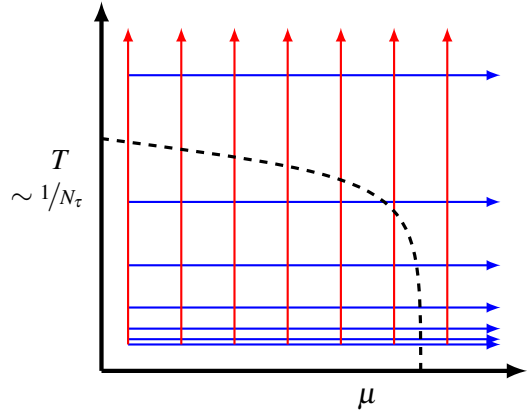
\begin{figure}[h!]
\centering
\begin{tikzpicture}[scale=1.4]
\colorlet{darkred}{red!50!black}
\colorlet{darkblue}{blue!50!black}
\colorlet{darkgreen}{green!50!black}
\begin{scope}[>=latex]

\draw[blue,->, thick] (0.25,0.25) -- (3.75,0.25); 
\draw[blue,->, thick] (0.25,0.30) -- (3.75,0.30); 
\draw[blue,->, thick] (0.25,0.40) -- (3.75,0.40); 
\draw[blue,->, thick] (0.25,0.60) -- (3.75,0.60); 
\draw[blue,->, thick] (0.25,1.00) -- (3.75,1.00); 
\draw[blue,->, thick] (0.25,1.60) -- (3.75,1.60); 
\draw[blue,->, thick] (0.25,2.80) -- (3.75,2.80); 

\draw[red,->, thick] (0.25,0.25) -- (0.25,3.25); 
\draw[red,->, thick] (0.75,0.25) -- (0.75,3.25); 
\draw[red,->, thick] (1.25,0.25) -- (1.25,3.25); 
\draw[red,->, thick] (1.75,0.25) -- (1.75,3.25); 
\draw[red,->, thick] (2.25,0.25) -- (2.25,3.25); 
\draw[red,->, thick] (2.75,0.25) -- (2.75,3.25); 
\draw[red,->, thick] (3.25,0.25) -- (3.25,3.25); 

\draw[black,->, ultra thick] (0.0,0.0) -- (0.0,3.50); 
\draw[black,->, ultra thick] (0.0,0.0) -- (4.0,0.00); 
\draw[black,very thick, dashed]  (3.0,0.0) .. controls (3,1.80) and (3,1.80) ..
(0,2.2); \node[below=0.1cm] at (2.5,0) {$\mu$};
\node[left=0.1cm] at (-0.12,2) {$T$};
\node[left=0.1cm] at (0,1.65) {$\sim \nicefrac{1}{N_\tau}$};
\end{scope}
\end{tikzpicture}
\caption{Strategy for determining the phase diagram of heavy dense QCD. The red
and blue lines indicate scans for fixed $\mu$ or fixed temperature $T$, at a
given $\beta$ and $\kappa$.}
\label{Bild3}
\end{figure}

We perform complex Langevin simulations for fixed gauge coupling, volume
and hopping parameter, more precisely we use the following setup: 
\begin{itemize}
\item Volume: $8^3 \times N_\tau$ 
\item $N_f=2$
\item $\beta = 5.8$
\item $\kappa = 0.04$ 
\item $N_\tau =
2,\;3,\;4,\;5,\;6,\;7,\;8,\;9,\;10,\;12,\;14,\;15,\;16,\;18,\;20,\;24,\;28$
\item $\mu = 0.0 - 3.6$ 
\end{itemize}
The corresponding lattice spacing $a \sim 0.15\,\mathrm{fm}$ has been
determined using the gradient flow~[12,13]. In the heavy dense theory the quark
mass is given by
\begin{equation}
 m_q \equiv - \ln(2 \kappa) = 2.53 \sim \mu_c,
\end{equation}
which also corresponds to the critical chemical potential $\mu_c$, i.e. the
chemical potential where the onset occurs at zero temperature.
We vary the chemical potential for fixed temporal extent $N_\tau$, or
equivalently fixed temperature $T$, to determine the transition to high densities by studying the fermion density,
\begin{equation}
n = \frac{1}{N_\tau N_s^3} \frac{\partial \mathrm{ln}\, Z}{\partial \mu}.
\end{equation}
For the thermal deconfinement transition we simulate different temperatures for
a given chemical potential $\mu$ and look at the behaviour of the Polyakov loop
$P_{\vec{x}}$. A graphical illustration of this strategy is shown in
figure~\ref{Bild3}. In total we have simulated $1024$ different combinations of
$N_\tau$ and $\mu$. These simulations have been extended to at least $100$ units
in Langevin time, so that we have roughly $2000$ independent measurements taking
into account thermalisation and auto-correlation.

\section{Results}

\begin{figure}[h!]
	\centering
	\begin{minipage}{0.65\linewidth}
	\centering
	\includegraphics[width=\linewidth]{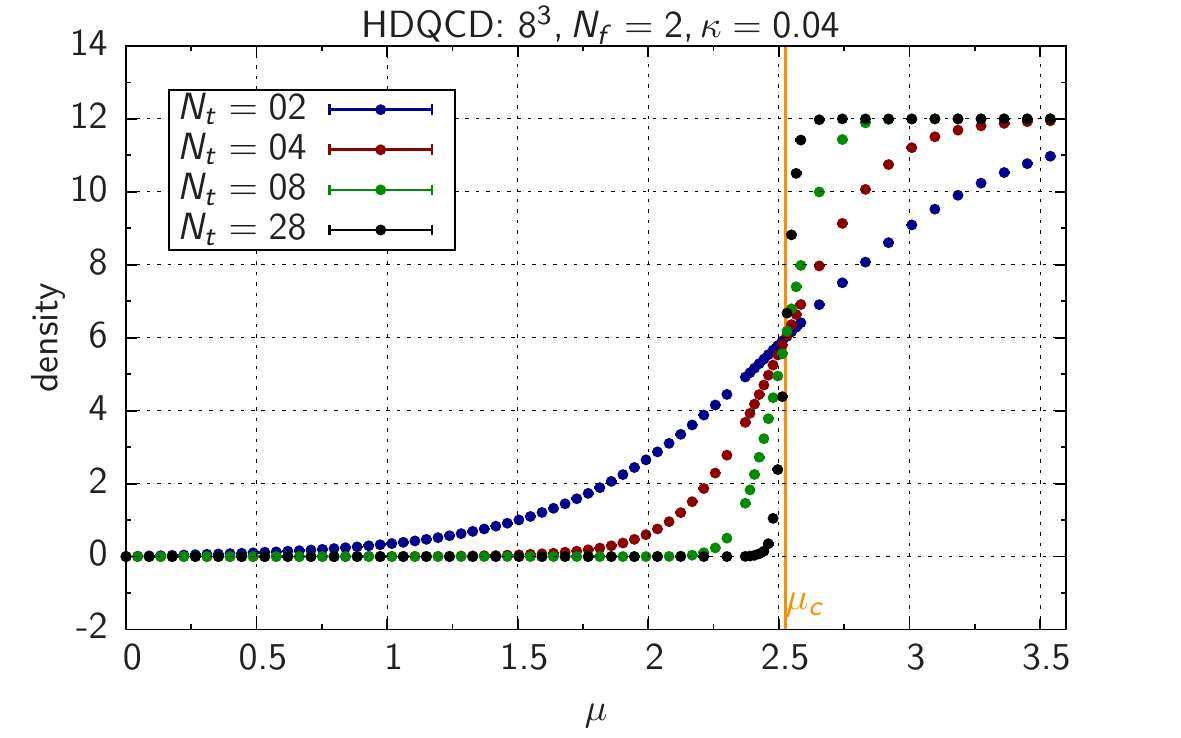}
	\end{minipage}
\caption{The fermion density as a function of the chemical potential $\mu$. The
expected critical chemical potention $\mu_c = m_q = - \ln(2 \kappa)$
is shown as well.}
	\label{Bild4}
\end{figure}
Figure~\ref{Bild4} shows the fermion density as a function of the chemical
potential for $4$ different temporal extents or temperatures.
The transition to high densities can be seen around the predicted value for the
critical chemical potential $\mu_c$. As expected the transition 
becomes sharper for lower temperatures. The saturation
to $n_\mathrm{sat} = 2\cdot N_c \cdot N_f = 12$ in figure~\ref{Bild4} is a pure
lattice artefact:
As soon as every lattice site has been filled with $12$ fermions, Pauli blocking prohibits more 
fermions to be added. 
\begin{figure}[h!]
	\centering
	\vspace{-0.5cm}
	\begin{minipage}{0.75\linewidth}
	\centering
	\includegraphics[width=\linewidth]{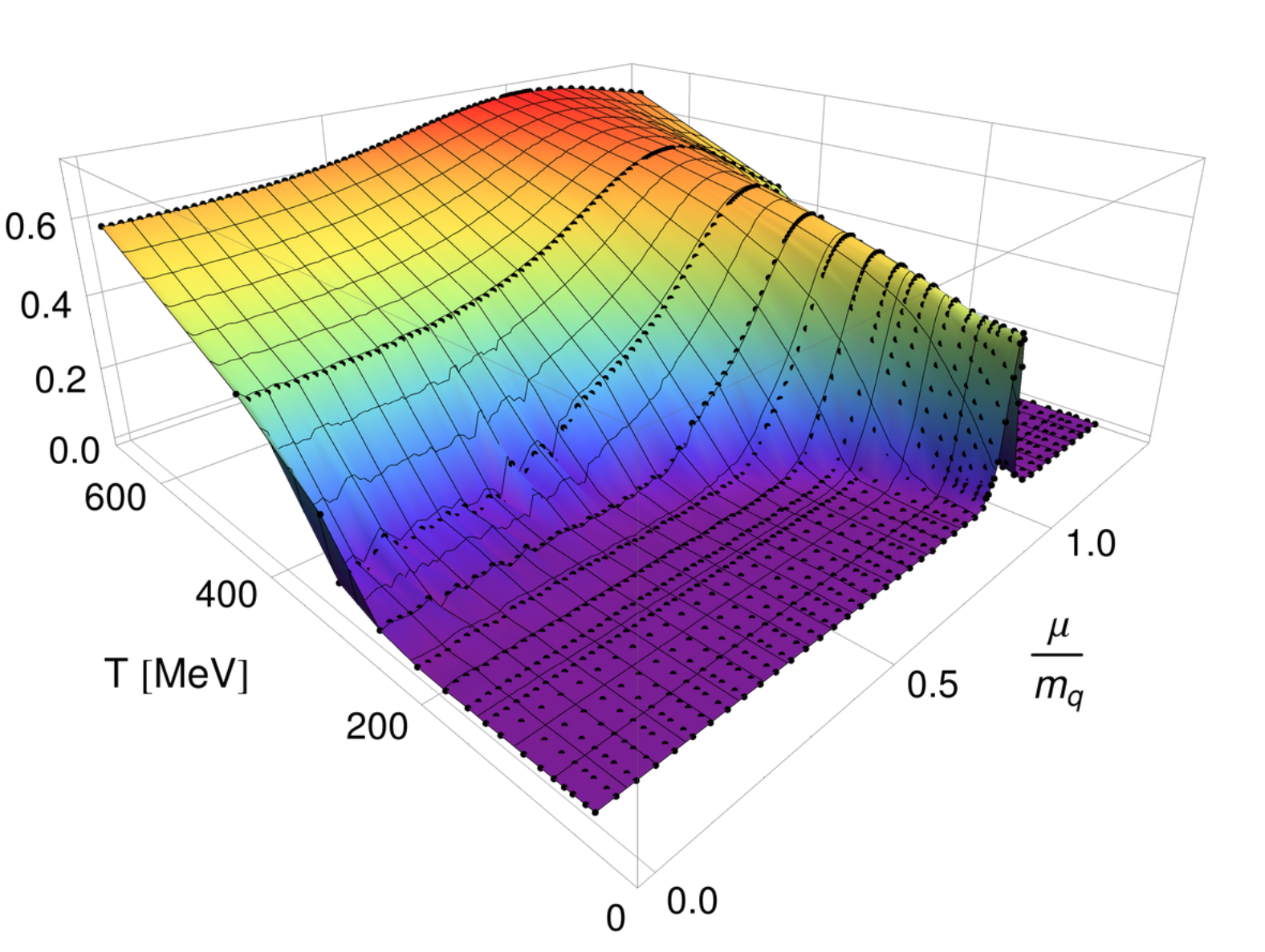}
	\end{minipage}
\caption{The real part of the Polyakov loop as function of the temperature and
chemical potential for fixed $\beta$ and $\kappa$. The black points are the
results of complex Langevin simulations for a given value of $\mu$ and $T$. 
The surface is the result of a cubic interpolation, in which the value
of the Polyakov loop is encoded in the colour.}
	\label{Bild5}
\end{figure}
Results for the Polyakov loop are shown in figure~\ref{Bild5} as a function
of $\mu$ and $T$. Each black point represents the results from one Langevin
simulation. The coloured surface is the result of a cubic interpolation to
estimate the intermediate behaviour, since the resolution
in temperature is limited due to the discrete steps for the temporal extent. 
As the temperature is increased, the Polyakov loop shows the thermal
deconfinement transition at fixed $\mu$. At fixed $T$, increasing $\mu$ leads to
the transition to the high-density phase. The subsequent drop of the Polyakov
loop, beyond $\nicefrac{\mu}{m_q} \simeq 1$, is a lattice artefact due to
saturation.

\begin{figure}[h!]
	\centering 
	\hspace{-0.0cm}
	\begin{minipage}{0.88\linewidth}
	\centering
	\includegraphics[width=\linewidth]{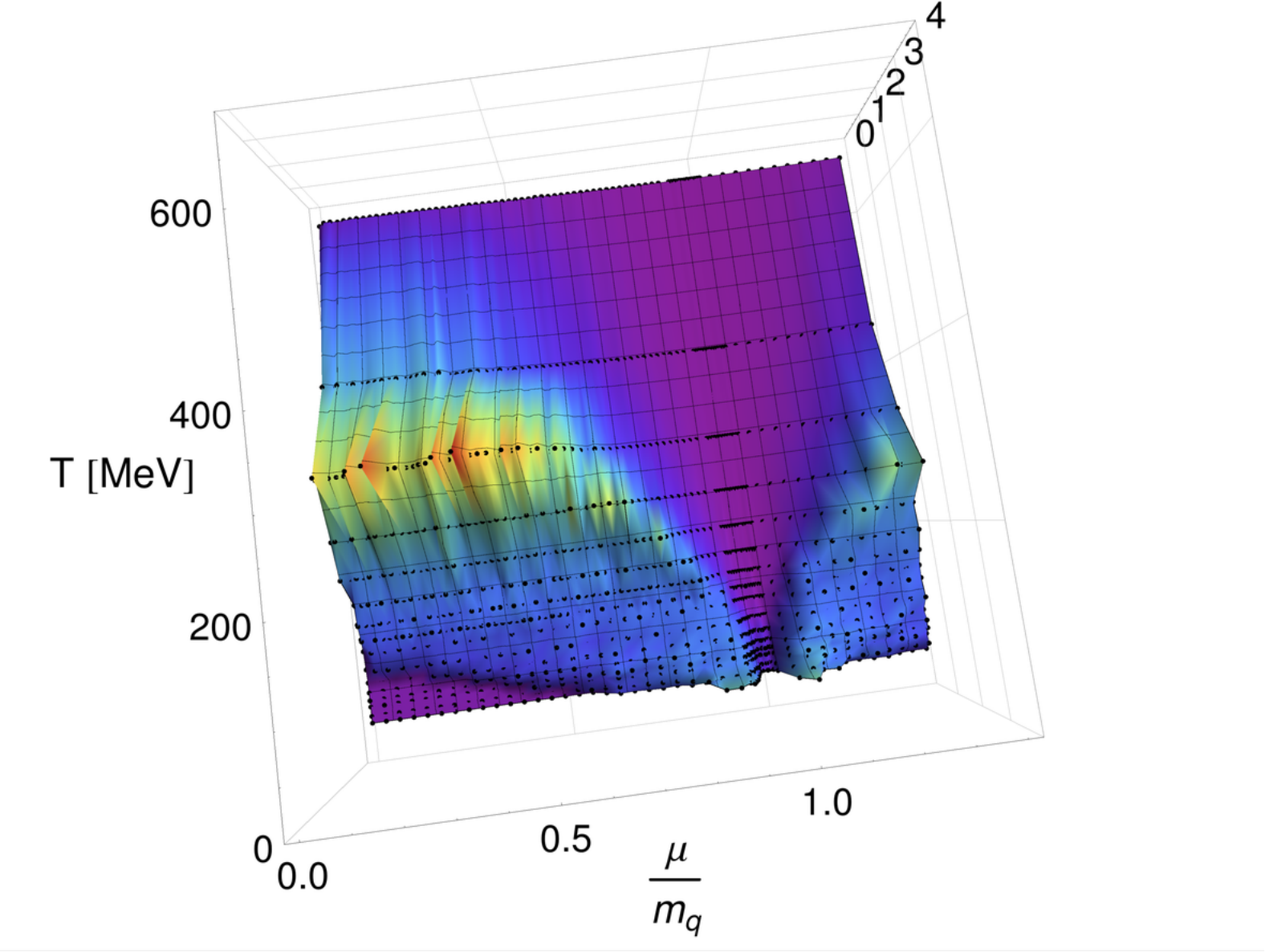}
	\end{minipage}
\caption{The Polyakov loop susceptibility as function of the temperature and
chemical potential for fixed $\beta$. The black points are the results of 
complex Langevin simulations for a given value of $\mu$ and $T$.
The surface is the result of a cubic interpolation, in which the value
of the Polyakov loop susceptibility is encoded in the colour.}
	\label{Bild6}
\end{figure}

To identify the order of the transition we have
started to study susceptibilities of these observables. The result
for the Polyakov loop susceptibility is shown in figure~\ref{Bild6}.
The boundary of the phase diagram is clearly visible. 
At $\nicefrac{\mu}{m_q} \simeq 1$ the Polyakov loop has a maximum and the
susceptibility vanishes. However, as mentioned above, this is a lattice artefact.
On the other hand, the peak in the susceptibility at lower $\mu$-values is
physical and indicates the transition to the deconfined phase. 
Intriguingly, at very low temperature and chemical potential there is an
indication that the Polyakov loop and its susceptibility increase from
zero while still in the confined phase. This requires further study. More
simulations with different gauge couplings will allow us to study the intermediate region and
improve the resolution for higher temperatures. By varying the volume 
and computing Binder cumulants~[14] we will be able to determine the
order of the phase transition as $\mu$ and $T$ are varied.

\section{Conclusions and Outlook}

Lattice simulations using the complex Langevin method deliver a reliable method
to determine the phase diagram for QCD in the limit of heavy quarks. Future
improvements include the variation of the simulation volume and the gauge
coupling in order to determine the phase diagram more accurately as well as the 
order of the phase transition. The study of heavy dense QCD serves both as
a guideline for future studies of "full" QCD~[12] including dynamical fermions,
as well as a proof of principle. 
\newline \phantom{bla}\newline

{\bf Acknowledgments:} 
We are grateful for the computing resources made available by HPC
Wales and by STFC through DiRAC computing facilities. This work is
supported by STFC, the Royal Society, the Wolfson Foundation and the Leverhulme Trust. 
FA is grateful for the support through
the Brazilian government program "Science without Borders" under scholarship number Bex 9463/13-5.

\end{document}